\documentstyle[prl,eqsecnum,aps]{revtex}
\begin{document}

\title{
WIGNER CRYSTALIZATION IN THE LOWEST LANDAU LEVEL FOR $\nu \ge 1/5$}

\author {
 V. A.\ Kashurnikov }
\address{
Moscow State Engineering Physics Institute,
115409 Moscow, Russia }

\author{N. V. Prokof'ev, B. V. Svistunov and I. S. Tupitsyn}
\address{
Russian Research Center "Kurchatov Institute", Moscow 123182, Russia }

\maketitle

\begin{abstract}
By means of exact diagonalization we study the low-energy states 
of seven electrons in the lowest Landau level which are confined
by a cylindric external potential modelling the rest of a 
macroscopic system and thus controlling the filling factor $\nu $.
Wigner crystal is found to be the ground state for filling factors between 
$ \nu = 1/3$ and  $ \nu = 1/5$ provided electrons interact via the
bare Coulomb potential.
Even at $\nu =1/5$ the solid state has lower energy than the 
Laughlin's one, although the two energies are rather close. 
We also discuss the role of pseudopotential parameters in 
the lowest Landau level and demonstrate that the earlier reported gapless
state, appearing when the short-range part of the interaction is suppressed,
has nothing in common with the Wigner crystalization in pure Coulomb case.
\end{abstract}

\bigskip

\section{Introduction}
\label{sec:1}

After the Laughlin states (LS) were proposed as new ground states 
of   strongly correlated  2D-electron liquid  in external
magnetic field \cite{1}, they were intensively compared to the
known ground states (GS), in particular with the Wigner crystal  (WC),
to understand the conditions and limitations of the experimental observation 
of the fractional quantum Hall effect (FQHE). In fact, even LS
themselves may be called "liquid" only for sufficiently large
filling factors $\nu = 1/m $ (for $m \le m_c \approx 71$ \cite{2}),
as follows from the formal analogy between the LS and the two-dimensional
one-component plasma at dimensionless temperature $T=1/2m$.
At very small $T$ the equivalent plasma undergoes 
Kosterlits-Thouless transition
to the state with the finite shear modulus and should be rather viewed
as a solid \cite{3}. 

For the Coulomb system, however, the critical value, $m_c$, is 
of academic importance only, since 
it is easy to prove that in the solid phase the Lauglin state 
differs qualitatively from the genuine GS.
Indeed, in a solid with non-zero shear modulus
and Coulomb interaction between the particles the sound dispersion  
law in magnetic field is $\omega_k \sim k^{3/2}$ \cite{4}.
Calculating the mean-square displacement  in such a solid 
at $T=0$ one finds a convergent answer 
$<(u(0)-u(R \to \infty ))^2> \sim \int d^2k/\omega_k \to const $.
On another hand, the equivalence between the LS and the finite-temperature
2D plasma implies divergency of this correlator. [In solid 
plasma this divergency is due to the transverse-sound dispersion law:
$<(u(0)-u(R \to \infty ))^2> \sim T \int d^2k/\omega_k^2 \sim T ln(R)$].
Thus solid LS maintains  the topological order only, which is
typical for the 2D solid with
short-range interactions in magnetic field when $\omega_k \sim k^2$,
in agreement with the fact that LS is a perfect
trial function for GS of the system with short-range interactions.

Having established that in the Coulomb system LS provides incorrect GS
for large $m>m_c$, one may further suspect that it may give the way
to the Wigner crystal at much smaller $m$. 
Early variational calculations performed for the 
electrons at the lowest Landau level and neglecting Landau-level-mixing 
effects \cite{5} gave strong evidence that WC has lower energy already at
$\nu =1/7$, which explained why there is no Hall conductivity quantisation 
at this filling factor. Mixing effects, which are very important 
in real systems, were taken into consideration in Refs.\cite{6}. 
It was found that virtual transitions between the Landau levels 
promote WC states and make LS unstable even at $\nu =1/3$ 
for sufficiently large mixing parameter $\lambda = 
\nu^{1/2}(e^2/\epsilon l_H)/\omega_c$, where $l_H=(1/eB)^{1/2}$
is the magnetic length in the external field $B$, $\epsilon$ is the dielectric
constant, and $\omega_c = eB/m^*$ is the cyclotron frequency for
electrons with the effective mass $m^*$. (We use units 
$\hbar = c=1$). For the $\lambda =0$ case the results
of Refs.\cite{5,6} predict LS to be the ground state for filling factors $\nu =1/3$ and
$1/5$.

An essential drawback of previous calculations is their variational
character. Since the difference in energy (per particle) between the 
LS and WC is only a few percent in Coulomb units 
$e^2/\epsilon l_H$, only filling factors $\nu =1/m$ with odd $m$,
were the liquid GS is known reasonably well, were analyzed
in detail. We simply do not know other liquid states with necessary
accuracy to compare them with the variational
functions we have at hand for WC. Moreover, it is impossible to
use LS energies $E_{LS}(1/m)$ to derive $E_{Liquid}(\nu )$ for other filling
factors by extrapolation, because of the cusps that must occur
at filling factors where FQHE exists. 

It is believed that due to this cusping down at simple rational $\nu$
there may arise reentrant WC-LS-WC behaviour around $\nu =1/5$
(and possibly $\nu =1/3$). Available experimental data seem to give
strong evidence that WC exists at $\nu =0.21$ \cite{7,8,9}.
However, this conclusion is based entirely on the divergent, activation-type 
resistivity $\rho_{xx} \to \infty $ at low temperatures, thus
one may not rule out the possibility of explaining the 
data by impurity induced electron localisation. 

The other way to study this problem is by means of exact diagonalization.
We are not aware of any systematic attempt to 
look for the WC states in numerical simulations of FQHE, although
in \cite{11} WC was suggested as a possibility to explain
perculiar degeneracies in the numerical spectra as a function of 
inter-particle pseudo-potential. As we show below, the collapse of the
LS  to the gapless state at $\nu =1/3$ found in \cite{11}, 
has nothing to do with the transition to WC state.
The best numerical calculations were done on a sphere for as many as
$N=12$ particles on 25 orbits \cite{10}. The spherical geometry being
perfect for the study of FQHE itself, is practically ineligible for
looking at Wigner crystallization. There is little doubt that the 
optimal electron configuration in the WC is the triangular lattice,
which is topologically prohibited on a sphere.

To give further theoretical support to the idea that WC state may be 
GS of the 2D electron gas at $\nu <1/5$, we calculated numerically
GS and low-laying excited states for  the cluster of 
7 electrons in the lowest Landau level and  in the confining potential.
(The number of particles in the system must be 7,12 etc., 
depending on the sample geometry, 
to account for the hexagonal symmetry of WC).
The confining potential was derived from the Coulomb interaction between an electron  
in the cluster and electrons outside the "first coordination
sphere" (that is at a distance equal or larger then $\sqrt{3}a$, where
$a$ is the atomic length in the triangular lattice). 
In fact, the radius, $a$, of this confining potential 
was our main variable determining an effective filling factor 
($a \sim \nu^{-1/2}$).
It was found that WC state (see below the discussion of what has to be 
thought of as WC for the system of only 7 particles) is GS
of the system for all filling factors between $0.4$ and $1/5$ except
for the region $0.34 > \nu > 0.294$ (including $\nu = 1/3$), where 
LS was essentially present in the structure of GS.
These results strongly suggest (keeping reservations for 
possible finite-size corrections) that LS
at $\nu =1/3$ is very close in energy to WC, and that  
in a perfect Coulomb system  at $\nu =1/5$ the ground state may be WC.
There is almost no doubt then that WC must exist between these two
filling factors.

Our calculation completely ignores impurities and Landau-level-mixing
effects, as well as possible screening of the bare Coulomb potential,
and the role of electron delocalisation in the direction perpendicular to 
the 2D plane. Some of these factors
are very important in real systems, and, e.g.,   electron screening, 
may work in favour of LS. We believe that experimental data of Refs. 
\cite{7,8,9}, demonstrating FQHE at $\nu =1/5$, 
can be accounted for along these lines.
We are planning to study different pseudo-potentials in a separate paper.

\section{Hamiltonian and numerical method}
\label{sec:2}
 
We study a 2D system of $N_e= 7$ electrons  
in the magnetic field in a confining potential.
Since we intend to model the macroscopic system
in the finite-cluster calculation, we derive this confining
potential as resulting from the Coulomb interaction between an 
electron in the cluster and other electrons outside the "first
coordination sphere". The unit length $a(\nu )$ in the WC lattice is 
expressed through the electron density $n_e$ and  filling factor 
$\nu =2\pi l_H^2 n_e$ as 
\begin{equation}
{\sqrt{3} \over 2}a^2 n_e =1 \;\;\mbox{or}\;\;a^2={4\pi \over \nu \sqrt{3}}\;,
\label{1}
\end{equation}
(from now on we measure all distances in units of $l_H$, which is 
kept fixed in our calculation). To account for the short-range correlations
we place surrounding electrons on the coordination 
spheres of the WC state. One may better view our system as  originating
from the  classical crystal with atoms sitting on the $m=0$
orbits $\Psi_j (z_j)=\Psi_{m=0} (z_j-Z_j)$, with the positions $Z_j$ forming
an ideal triangular lattice corresponding to the filling factor $\nu$. 
We then allow for full quantum dynamics of 7 electrons with $Z_j=0$ and
$\mid Z_j \mid =a$ while keeping other electrons frozen, but coupled to 
the first 7 ones by Coulomb forces. Thus obtained confining potential
is not spherically symmetric, and may cause transitions changing
the momentum of an inner electron by $\Delta \! m= \pm 6n $. 
Obviously, this 
coupling will promote "crystal" state for the central cluster. To 
avoid this shortcoming we ignore all these transitions, keeping only
the diagonal part of the interaction; this procedure is 
equivalent to rotational averaging of the confining potential.

We also ignore Landau level mixing, which means that Coulomb
interaction is the only energy scale in the problem. To simplify
the notation we measure all energies in units $e^2/(\epsilon l_H)$.
Working in the symmetric gauge 
$\vec{A} =1/2 H (-y,x)$ we place the electrons 
on up to 31 orbits corresponding to the angular momentum states ranging from
$m=0$ to $m=30$. 
The maximum number of orbits in a given
calculation was defined by the
condition that occupation numbers $<n(m)>$, giving the probability
to find an electron in the state with the orbital momentum $m$,
stop changing (at the level of 0.001), when the number of orbits 
is increased, and that the lagest momentum state be empty with the same
accuracy. 

The starting Hamiltonian then can be written as
\begin{equation}
H=\sum_{m_1,m_2,m_3,m_4} V_{m_1,m_2,m_3,m_4}a^{\dag}_{m_1}
a^{\dag}_{m_2}a_{m_3}a_{m_4}
+ \sum_m V_m^{(MF)} a^{\dag}_{m}a_{m}\;,
\label{2}
\end{equation}
where $a^{\dag}_{m}$  creates an electron in the state 
\begin{equation}
\mid m \rangle = {z^m \over \sqrt{2\pi 2^m m!}} e^{-\mid z \mid ^2/4} \;.
\label{3}
\end{equation}
In the symmetric gauge the interaction matrix elements in the first term 
in (\ref{2}) satisfy the conservation law $m_1+m_2-m_3-m_4 =0$ and may be 
written as
\begin{eqnarray}
V_{m_1,m_2,m_3,m_4}=&&{(-1)^{m_1+m_3} \over  2^{M}
( \prod_{i=1}^4  m_i! ) ^{1/2} }\:
\sum_{k_1=0}^{\le m_1,m_4}\: \sum_{k_2=0}^{\le m_2,m_3} \nonumber \\
&&\times (-2)^{k_1+k_2}\:  k_1!k_2!\:
C^{m_1}_{k_1}C^{m_4}_{k_1}C^{m_2}_{k_2}C^{m_3}_{k_2}
I(M,k_1,k_2)\;,
\label{4}
\end{eqnarray}
\begin{equation}
I(M,k_1,k_2)=\int_0^{\infty} dq {V_C(q)\over 2\pi }
\: q^{2M-2k_1-2k_2+1}e^{-q^2}     \;,
\label{5}
\end{equation}
where $M=m_1+m_2$, $C^{i}_{j}=i!/(j!(i-j)!)$, 
and $V_C(q)$ gives the 
Fourier component of the pair potential.
For the case of Coulomb interaction $V_C(q)=2\pi/q$, and 
the final expression simplifies to
\begin{eqnarray}
V_{m_1,m_2,m_3,m_4} =&& { (-1)^{m_1+m_3} \sqrt{\pi} \over 2^{2M+1} 
  ( \prod_{i=1}^4  m_i! ) ^{1/2} }\:
\sum_{k_1=0}^{\le m_1,m_4}\: \sum_{k_2=0}^{\le m_2,m_3} \nonumber \\ 
& &\times (-4)^{k_1+k_2}\: [2(M-k_1-k_2)-1]!! \:
C^{m_1}_{k_1}C^{m_4}_{k_1}C^{m_2}_{k_2}C^{m_3}_{k_2} \;.
\label{6}
\end{eqnarray}

To construct the confining potential one has to calculate the diagonal matrix
elements for one electron staying on orbit $\mid m \rangle$  
and the other electron staying on orbit $\Psi_{m=0}(z-Z_j)$
(for the diagonal matrix element   it does not matter whether the state
$\Psi_{m=0}(z-Z_j)$ is defined in the same gauge as states $\mid m \rangle$
or obtained by gauge transforming the state $\mid m=0 \rangle$). 
Thus we have
\begin{equation}
V_m^{(MF)} =\sum_{j=2,3,\dots }  V_m(Z_j)\;,
\label{7}
\end{equation}
where the sum is over all coordination spheres 
starting from the second one, and 
\begin{equation}
V_m(Z_j)=(-1)^{m}2^{-m}m!
\sum_{k=0}^{m}{(-1)^{k} 2^{k} \over k! [(m-k)!]^2}
\int_0^{\infty} dq {V_C(q)\over 2\pi }
\: q^{2m-2k+1}e^{-q^2}J_0(q \mid Z_j \mid ) \;.
\label{8}
\end{equation}
Here $J_0(q)$ is the Bessel function. In practice we constructed the
confining potential by summing over all coordination spheres 
inside the radius $100 l_H$.

Our diagonalization procedure is arranged as follows.
For the calculation of the groundstate level we use the standard
modified Lanczos method with the straightforward iteration procedure
(see, for example, \cite{12a}), while for the calculation of the lowest 
excited levels we apply more sophisticated method \cite{KPS}.
The set of approximate eigenfunctions is reconstructed 
from Relay's tridiagonal matrix \cite{Pis}, and the trial wavefunction is 
expanded in it. As is known, the set inevitably involves a substantial number 
of spurious states, due to numerical errors. These states, however, may be 
easily identified by their negligible contribution to the expansion of the 
trial wavefunction. Upon exclusion of the spurious states the set is 
subjected to the orthogonalization and correction by Newton method. The 
relative (with respect to a characteristic interlevel spacing) errors in the 
energy level calculation are typically of order 
$10^{-13} \div 10^{-11}$ for the 
groundstate, and of order $10^{-9} \div 10^{-5}$ for some ten first 
excited states. 
Since the Hamiltonian conserves the total angular momentum 
we take advantage of this symmetry to proceed separately for 
each angular momentum sector. 

\section{Results for the low-energy states}
\label{sec:3}

Before presenting our numerical results for the ground and first exited
states in the cluster, let us first discuss
how one may discriminate between the liquid and solid states in such a small
system. The most obvious solution is to look at the pair
correlation function. From the symmerty considerations we expect
(at least for small filling factors)
that one particle will be always staying near the origin, and the 
rest 6 particles will have their density distribution being peaked 
at a distance $\approx a(\nu )$ apart. 
These particles are mutually correlated over the angle $\theta $ between
their coordinates on the "first coordination sphere".
The appropriate pair correlation fuction thus can be defined as
\begin{equation}
g_{\alpha }(\theta )=\langle \alpha \mid \Psi^{\dag}(z_1) \Psi^{\dag}(z_2)   
\Psi (z_2) \Psi (z_1) \mid \alpha \rangle \;,
\hspace{1.5cm} \mid z_1 \mid =\mid z_2 \mid = a \;, 
\label{9}
\end{equation}
where $\theta = \arg (z_1)-\arg (z_2)$ varies in the interval $(0,\pi )$.
In the solid state we expect three well-defined oscillations in 
$g( \theta )$, while in the liquid these oscillations should be
strongly damped. It is difficult to predict apriori the amplitudes
of oscillations, but it is known (see, e.g., \cite{3}) that pair correlations
in the LS disappear very rapidly at $\nu =1/3$ and $\nu =1/5$. 
Our definition of $g$ is not 
quite standard, but we believe that its  qualitative behaviour is the same 
(we verify this point explicitly below). Anyway,
the abrupt change of the ground state correlator $g_{G}(\theta )$ 
as a function of $\nu $ is indicative of the solid-liquid transition.

One may also expect some qualitative differences in the 
structure of the low-energy spectra  of WC and LS. By construction,
our Hamiltonian is cylindrically symmetric and conserves the total
angular momentum $M =\sum_m m n(m) $. In the solid phase of the 
macroscopic system this symmetry is spontaneously broken. For the
triangular lattice under study the symmetry is broken by
coupling momenta $M_G \pm 6n$ (where $M_G$ is the ground state
angular momentum and $n$ is an integer). Thus in the solid phase we
expect the states $\mid M_G \pm 6n>$ to form a subset of the lowest
excited states well separated from the rest of the  excitation spectrum 
in these sectors. There is no special reason to have the lowest
excitations at $M_G \pm 6n$ in a liquid phase, nor should they
have much lower energies than excited states with $M=M_G$.

One remark is in order here. In a really macroscopic solid, the 
lowest states are those corresponding to the system rotation as a whole, 
with the energy going as $E \sim (M-M_G)^2/L^4$ where $L$ is the system's
size. The crystal symmetry is not present in the structure of the spectrum
explicitly, but it is important that the states, mixed by the symmetry
breaking fields, are among the lowest ones. In the finite system of only
seven particles we do not expect the spectrum to be quadratic in
$M-M_G$, since this property in the rotating solid
is achieved by creating extra zeros in the wave-function
$\tilde{\Psi }(z_j) = \Psi (z_1,z_2,\dots ,z_N )$ for fixed 
$\{z_1, \dots,z_{j-1},z_{j+1},\dots ,z_N \}$. This procedure may be 
too costy in energy in a small system. 
For $N_e=7$ rotation is equivalent to the 
correlated motion of six particles. The first rotating state which 
requires no extra zeros in  $\tilde{\Psi }(z_j)$ is that with $\mid M-M_G \mid =6$.

The other point concerns the consistency of our 
procedure of controlling the filling factor according to 
equations (\ref{1}) and (\ref{8}). 
Since the confining potential is derived from the crystal state, 
a natural question arises of how good is this 
approximation for modelling a liquid environment. 
There is no doubt that at $\nu =1/3$ the ground state
is well described by LS with  the ground-state angular momentum 
$M_G =3 N_e(N_e-1)/2 =63 $.
No matter how trivial, this fact is not at all predetermined
by the numerical procedure used. Its validity was confirmed in our
calculations, thus demonstrating consistency between Eq.~(\ref{1})
and the effective filling factor.
Similarly, we observed that 
$M_G = 5 N_e(N_e-1)/2=105$ when $\nu =0.198 $ in Eq.~(\ref{1}).
The consistency of our "mean-field" procedure follows also from
the fact that for all $\nu <1/2$ the position of the maximum in the 
particle density $\rho ( R)$ coinsides with $a(\nu )$.

In Table I we present our data for the ground state angular momentum
as a function of filling factor. For $\nu >0.705$ the system
is described by the IQHE state with occupation numbers
$n_i=1$ for $i=0,1,\dots ,6$. After drastic transformations in
the range of densities between $0.705$ and $0.46$ GS
 evolvs into the state with well-defined pair 
correlation function $g(\theta )$. We note, that starting from 
rather high filling factor $0.587$ the angular momentum
of the ground state changes by $6$. 
Also, the lowest exited state is always in the sector $M_G \pm 6$.

To identify the nature of GS we present in
Fig.1 the plots of $g(\theta )$ calculated for critical filling factors 
$\nu_M$ where $M_G$ jumps. While going from $M_G=45$ to $M_G=51 \to 57 $ the
pair correlation function  develops more pronounced 
oscillations. We naturally consider this evolution
as formation of more rigid solid state order in the system,
although the filling factor seems to be too large here to expect WC state
in a macroscopic system. If we ignore for the moment what is happening
in $M_G=63$ then the "crystal set" may be smoothly continued to 
higher momentum states $57 \to 69 \to 75 \to 81 \dots \to 111$
resulting finally in a quite impressive "long-range" order (see Fig.2).
With all the reservations concerning small system size we have to conclude 
that WC has lower energy than LS in the range of filling
factors between $1/3$ and $1/5$. 

We also observe a well-defined structure
of "satellite states" $\mid M_G \pm 6n>$ in the energy spectrum 
for small $\nu $, for example, when $M_G=81$ we find that 
 $E_{75}-E_G$ and $E_{87}-E_G$ are some
five times smaller than the energy of the first excited state 
in sectors $M=75,81,87$ (see Fig.3). Note also the remarkable 
similarity between  the low-energy specra in the basic set 
of states  with $M=M_G \pm 6n$.

Clearly, the state with $M_G=63$ is special in that its pair 
correlation function is more "liquid-like" than $g(\theta )$ for both 
$M_G=57$ and $M_G=69$.  As mentioned above LS at $\nu =1/3$ has 
$M=63$, thus irregular behaviour of the pair correlation function
in this sector may be due to the change of GS from
solid to liquid. This suggestion seems to be correct, because
the calculated projection of the exact GS
for $\nu =0.32$, i.e., in the middle of the stability interval
of the sector $M=63$ (see Table I),
on the Laughlin state is as large as 
$<\Psi_{LS}^{(1/3)} \mid \Psi_G^{(63)}>=0.934$,
and the ground state energy is extremely well approximated 
by the variational value $E_{LS}^{(1/3)}=<\Psi_{LS}^{(1/3)} \mid  
H \mid \Psi_{LS}^{(1/3)}>$. In Coulomb units we find $E_{LS}^{(1/3)} -
E_{G}^{(63)}= 0.0134$, while the energy of the first excited  state in
the sector $M_G=63$ is almost five times higher, $E_{1}^{(63)}-
E_{G}^{(63)} =0.0621$. 
Furthermore, there is no pronounced satellite structure in
the low-energy spectrum when $M_G=63$.
Surprisingly enough, the ground state wave
function and $g_G(\theta )$ are rather different from  
$\mid \Psi_{LS}^{(1/3)}>$ and $g_{LS}(\theta )$. It is clearly seen
in Fig.4 that $g_{LS}(\theta )$ is almost flat for large $\theta $ and
shows no sign of pair correlations across the diameter of our system. 
These correlations are present in GS. Also, in Fig.5
we plot the average occupation numbers $<n(m)>$, calculated in GS
and in LS. We see that $<n(m)>$ in LS has much smaller amplitude at $m=0$
and more shallow minimum. As one might expect beforhead, the
central particle is not at all localized in the liquid phase.

To clarify the nature of such differences, we construct another
variational state, which may be regarded as solid, 
$\mid \tilde{\Psi}^{(63)}>$. Consider two nearest solid states,
e.g., $\mid \Psi_{G}^{(75)}>$ and  $\mid \Psi ^{(81)}>$
at some $0.255<\nu <0.276$.
We notice that their distribution functions $<n(m)>$ are very close
in shape (see Fig.6), with one particle being localised on orbits
with small $m$ (actually $m=0,1$; the sum of $<n(m)>$ before the
minimum is almost $1$), and the other six particles occupying 
extended states with large $m$. When going from $M_G=75$ to 
$M=81$ the value of $M=\sum_m m<n(m)>$ changes by $6$ almost entirely
due to the change of the occupation numbers of six particles on the 
first coordination sphere, i.e., $<n^{(81)}(m+1)> \approx 
<n^{(75)}(m)>$ for large $m$. Considering $\mid \Psi ^{(81)}>$ as rotating 
state with all the pair correlations being preserved, we may construct
the variational state $\mid \tilde{\Psi}^{(75)}>$ close to 
exact $\mid \Psi_G ^{(75)}>$ according to the rule
\begin{equation}
\mid \tilde{\Psi}^{(75)}> \sim \sum_{\{ m_i \}} C^{(81)}_{\{ m_i \}}
a^{\dag }_{m_7-1}a^{\dag }_{m_6-1} \dots a^{\dag }_{m_2-1}a^{\dag }_{m_1}
\mid 0>\;,\;\;\;\;(m_{i+1}>m_i)\;,
\label{10}
\end{equation}
where $\sum_i m_i =81$, and $C^{(81)}_{\{ m_i \}}$ are the corresponding
exact amplitudes of the expansion 
$\mid \Psi^{(81)}> = \sum_{\{ m_i \}} C^{(81)}_{\{ m_i \}}
\prod_i a^{\dag }_{m_i} \mid 0>$. Notice that the first particle
keeps its states. This procedure is well justified because the first 
particle is separated from the others by a deep minimum in the distribution 
function (with $<n(m)>$ close to zero in minimum, see Fig.6). To estimate
the accuracy of this procedure we project thus obtained variational state
on exact $\mid \Psi_G ^{(75)}>$ and find the overlap to be
$0.995$. We apply now this method to construct 
$\mid \tilde{\Psi}^{(63)}>$ from $\mid \Psi^{(69)}>$ obtained at $\nu =0.32$,
to obtain  a solid-state trial wave-function. In Fig.4 and Fig.5 we show
the pair correlation function and $<n(m)>$ of this  state.
Finally, the solid-state variational energy turns out to be as 
good as $E_S^{(63)}-E_G^{(63)}=0.0075$  and the overlap with GS is
$<\tilde{\Psi}^{(63)} \mid \Psi_G^{(63)}> = 0.953$
(even better than that of the Laughlin state!).
 
From these data we have to conclude that the genuine GS in 
the range of filling factors $0.294 <\nu < 0.340$ is a mixture
of solid and liquid phases with comparable amplitudes. Not only these 
two quite different states strongly overlap with the ground 
state and almost minimize the energy, but also 
$<\tilde{\Psi}^{(63)} \mid \Psi_{LS}^{(1/3)}> = 0.817$. That large
overlap is, of course, the finite-size effect. Obviously,
under these conditions no definite conclusion about the true GS of the 
macroscopic system is possible, and there is no contradiction with the
experimental fact that at $\nu =1/3$ the GS is the Laughlin 
liquid. 

We would like to comment here on the widly used argument, based on 
diagonalization of {\it finite-size} systems,
that large overlap of  LS with the exact GS and its precise 
energy, may serve as a criterion that the corresponding  
macroscopic system will be an incompressible liquid. We have demonstrated
above that this argument simply does not work for the system
of seven particles; short-range order in  LS and WC
turns out to be very similar. One has to analyze more delicate   properties (like 
pair correlation function at large distances) to discriminate between the
two phases.  

It follows from our data in Fig.2 that GS in the sector
$M=105$ is of solid type. To see how different is $\mid \Psi_{G}^{(105)}>$
from $\mid \Psi_{LS}^{(1/5)}>$ we present in Fig.7 the corresponding
correlation functions. We further confirm this result by
calculating the overlap between the two states, 
$<\Psi_{LS}^{(1/5)} \mid \Psi_G^{(105)}>=0.759$, and the Laughlin state energy
$E_{LS}^{(1/5)} -E_{G}^{(105)}= 0.0188$ (compare with the energy of the 
first excited state $E_{1}^{(105)}-E_{G}^{(105)} =0.0319$).
Now, the admixture of the LS in the structure of GS is much smaller than
that at $\nu =1/3$ and the variational energy is of the order of the first
excited state in this sector. 
To reconcile this result with the experimental
observation of the FQHE at $\nu =1/5$ in some (not all!) systems
\cite{7,8,9}, we notice that our result was obtained on finite-size
system and for the unscreened Coulomb interaction between the particles.
Given rather large difference in energy between LS and WC
found in our study, it is likely that WC will be the true 
GS of a macroscopic system too. This conclusion, however, 
may change for the screened Coulomb interaction since the
Laughlin state is stabilized by short-range interactions. We plan 
to investigate the role of screening effects on the ground state
at $\nu =1/5$ in a separate paper. 

Since the liquid energy is casping down at $\nu =1/5$, our results
give very strong support to the idea that WC exists in the Coulomb
system for $\nu >1/5 $. Even if WC is replaced with  LS at $\nu =1/5$
when the interaction potential is screened, it will most likely
survive at slightly larger filling factors. We thus conclude that
experiments \cite{7,8,9} did see WC state around $\nu =1/5$.

\section{Other ground states in the pseudo-potential approach}
\label{sec:4}

It was found in Ref.\cite{11} that varying pair potential between 
the particles in the lowest Landau level one can drastically
change the nature of the ground state. 
In this section we discuss whether this change
is of any relevance to Wigner crystallization.

Following Ref.\cite{11} we characterize the 
potential by the energies, $U_m$, of pairs of particles with relative 
angular momentum $m$. In the lowest Landau level
\begin{equation}
U_m=\int_0^\infty dq q \left( {V(q) \over 2\pi } \right) e^{-q^2} 
L_m(q^2) \;,
\label{11}
\end{equation}
where $L_m$ are the Laguerre polynomials. These are pseudopotential
parameters because different bare interactions may have the same
values of $U_m$. For the Coulomb interaction these parameters are 
$U_m = \sqrt{\pi } (2m-1)!!/(2^{m+1}m!)$ and decrease slowly
with $m$. Spinless fermions are coupled with odd values of $m$ only. 
The effect of decreasing $U_1$ for the Coulomb
system of $N_e=6$ electrons on a sphere  at $\nu=1/3$ was the 
collapse of the Laughlin-type ground 
state to some gapless state \cite{11} (we will call it $U_1$-state). 
The nature of this state was not clearly
identified, although the results did suggest a tendency to 
charge density wave formation. As we demonstrate below, the gapless ground 
state obtained by reducing the short-range part of the Coulomb interaction
{\it is not} the conventional Wigner crystal (by "conventional"
we mean the single-atom triangular lattice).

We start by noting that the new state has almost zero overlap with 
LS \cite{11}. This result is in sharp contrast with the large overlap between
WC and LS found in Sec. \ref{sec:3}. This fact alone is sufficient to rule
out WC as a candidate for the $U_1$-state. Furthermore, as is seen from 
the data presented in Ref.\cite{11},
the collapse of LS {\it is not} accompanied by formation of the low-energy 
satellite states corresponding to the rotations of the octahedron 
formed by six particles on a sphere. 

We performed an analogous study of the ground state changes as a 
function of the $U_1$ pseudopotential parameter for our system of 
seven particles. In agreement with Ref.\cite{11} we observe a drastic
transformation of the ground state at $\nu =0.32 $ when $U_1$
is reduced to $0.35$. For smaller values of $U_1$ the ground state
angular momentum changes from $M_G=63$ to $M_G=56$. The change of $M_G$
by 7, not by 6, also proves that we are 
not dealing with the conventional WC. Finally, we followed the transformation
of the solid ground state with $M_G=75$ at $\nu = 0.265$ and observed its 
collapse to the same $U_1$-state for $U_1 <0.32$. These results leave no
doubt that reducing the short-range part of the Coulomb potential
promotes new ground state other than LS or WC.

To have a better filling about real-space interaction potentials
with reduced values of $U_1$ we show in Fig.8 the particular set of
interaction potentials of the form
\begin{equation}
V(r)={ \sqrt{\pi } \over 2} e^{-r^2/8} I_o(r^2/8) 
-{\lambda \over 2}e^{-r^2/4}  \;,
\label{12}
\end{equation}
where $I_o$ is the Bessel function. The first term gives the Coulomb
interaction between the two unit charges at a distance 
$r=\vert r_1-r_2 \vert $ apart, each being spread out with the 
Gaussian distribution $(2\pi )^{-1/2}\exp \{-\mid z-r_i \mid ^2/2 \}$, and
the second term further suppresses the short-range part of the first. The
choice of $V(r)$ in this form is kind of arbitrary.
It is justified by the simplicity of its Fourier transform
$V(q)/2\pi = (1/q -\lambda )\exp \{ -q^2 \}$. In a more general case
one may also vary the "cutoff length" by letting $r \to r/r_c$ in 
the second term. In Fig.8
we plot the potential (\ref{12})  for $\lambda = 0.8$, $1.0$, and $1.2$.
The corresponding values of $U_m$ are given in Table II.
We see that $U_1$-state is stabilized at the edge of digging a potential
well at short distances.

In Fig.9 we present $<n(m)>$ distribution in the $U_1$-state
with momentum $M_G=56$. Quite unexpectedly  in the $U_1$-state, 
the central particle is replaced with the 
correlated hole. One has to appreciate this result in the system
with the long-ranged Coulomb potential - by taking the central particle from
orbits with $m=0,1$ and placing it to much higher orbits we 
substantially increase its mean-field energy. On another hand
the "first coordination sphere" of six particles moves to 
internal orbits thus gaining some mean-field energy. 
Thus we see, that $U_1$-state suggests locally ( on the scale of $a$)
inhomogeneous particle distribution. Of course, the long-range tail 
of the Coulomb potential ensures that the macroscopic system is homogeneous
on a large scale $\gg a$, but when the short-range part of the interaction 
is reduced, the system
may choose states with local density higher than average. 
We are not able to say anything definite about such a state except that
it is not conventional WC. Obviously, if the final state is a solid with
more than one particle in the unit cell, it can not be traced from the 
numeric study of seven particles.  
  
\section{Acknowledgement}
\label{sec:5}
We are grateful  to  A.I. Podlivaev  for his assistance in
writing the exact-diagonalization code.
This work was supported  by the Russian Foundation for
Basic Research (95-02-06191a). 

\section{Note Added}
After this work was completed we became aware
of the fact that we had overlooked some important experimental results
\cite{Dorozh} which seem to be in an excellent agreement with our 
numerical study. In these Refs.\ a metal - insulator transition is
found to occur at the universal filling factor $\nu_c \simeq 0.28$
in rather wide range of magnetic fields and sample mobilities, no 
reentrant behavior is observed around $\nu = 1/5$. The authors argue
that their results could be explained in terms of Wigner crystallization
[though other interpretations are not ruled out].

\begin{table}
\setdec 0.000
\caption{GROUND-STATE ANGULAR MOMENTUM}
\begin{tabular}{ccccccccccccccccc}
GS angular     &  &  &  &  &  &  &  &  &  &  &  &  &  &  &   &    \\
momentum  &  &  &  &  &  &  &  &  &  &  &  &  &  &  &   &    \\
$M_G$          &21    &   28&   33&   39&   45&   51&   57&   63&   69&   75&   81&   87&   93&   99&  105&111 \\
\tableline
Range of   &  &  &  &  &  &  &  &  &  &  &  &  &  &  &   &    \\ 
filling factors     &      &     &     &     &     &     &     &     &     &     &     &     &     &     &     &     \\
$\nu_{max}  $      & 1   &0.705&0.587&0.527&0.460&0.408&0.364&0.340&0.294&0.276&0.255&0.240&0.224&0.211&0.198&0.188 \\
$\nu_{min}  $      & 0.705&0.587&0.527&0.460&0.408&0.364&0.340&0.294&0.276&0.255&0.240&0.224&0.211&0.198&0.188&$\;$ \\
\end{tabular}
\label{t1}
\end{table}

\begin{table}
\setdec 0.000
\caption{PSEUDOPOTENTIAL PARAMETERS}
\begin{tabular}{lccc}
Potential      &  $U_1$  & $U_3$ &  $U_5$ \\
\tableline
Coulomb       &  $0.44 $   &  $0.28 $& $0.22 $  \\
$\lambda =0.8$ & $ 0.37$    & $ 0.29$ &$ 0.23$  \\
$\lambda =1.0$ & $ 0.35$    & $ 0.28$ &$ 0.23$   \\
$\lambda =1.2$ & $ 0.32$    & $ 0.27$ &$ 0.22$
\end{tabular}
\label{t2}
\end{table}

\figure{FIGURE 1. \\
Pair correlation functions $g(\theta )$ for the ground states
in the degeneracy points corresponding to the angular momentum
changes $51 \to 57$, $57 \to 63$, $63 \to 69$, and $69 \to 75$}
 
\figure{FIGURE 2. \\
Pair correlation functions $g(\theta )$ for the ground states
at the degeneracy points corresponding to the angular momentum
changes $87 \to 93$, $93 \to 99$, $99 \to 105$, and $105 \to 111$}
 
\figure{FIGURE 3. \\
Low-energy spectrum at $\nu =0.248$ ($M_G=81$) }

\figure{FIGURE 4. \\
Pair correlation functions $g(\theta )$ 
at $\nu =0.32$ for the ground state, the Laughlin state 
$\mid \Psi_{LS}^{(1/3)} >$, and
the solid state $\mid \tilde{\Psi }^{(63)} >$ }

\figure{FIGURE 5. \\
$<n(m)>$ distributions 
at $\nu =0.32 for the ground state$, the Laughlin state 
$\mid \Psi_{LS}^{(1/3)} >$, and
the solid state $\mid \tilde{\Psi }^{(63)} >$ }

\figure{FIGURE 6. \\
$<n(m)>$ distributions for the ground state $\mid \Psi_G^{(75)} >$
and the excited state $\mid \Psi^{(81)} >$ at $\nu =0.265$ }

\figure{FIGURE 7. \\
Pair correlation functions $g(\theta )$ 
at $\nu =0.193$ for the ground state and  the Laughlin state $\mid \Psi_{LS}^{(1/5)} >$ }

\figure{FIGURE 8. \\
Some realizations of the interaction potential in the real space
with reduced values of $U_1$.}

\figure{FIGURE 9. \\
$<n(m)>$ distribution  for the $U_1$-state at $\nu =0.32$
and $\lambda=1.2$. }

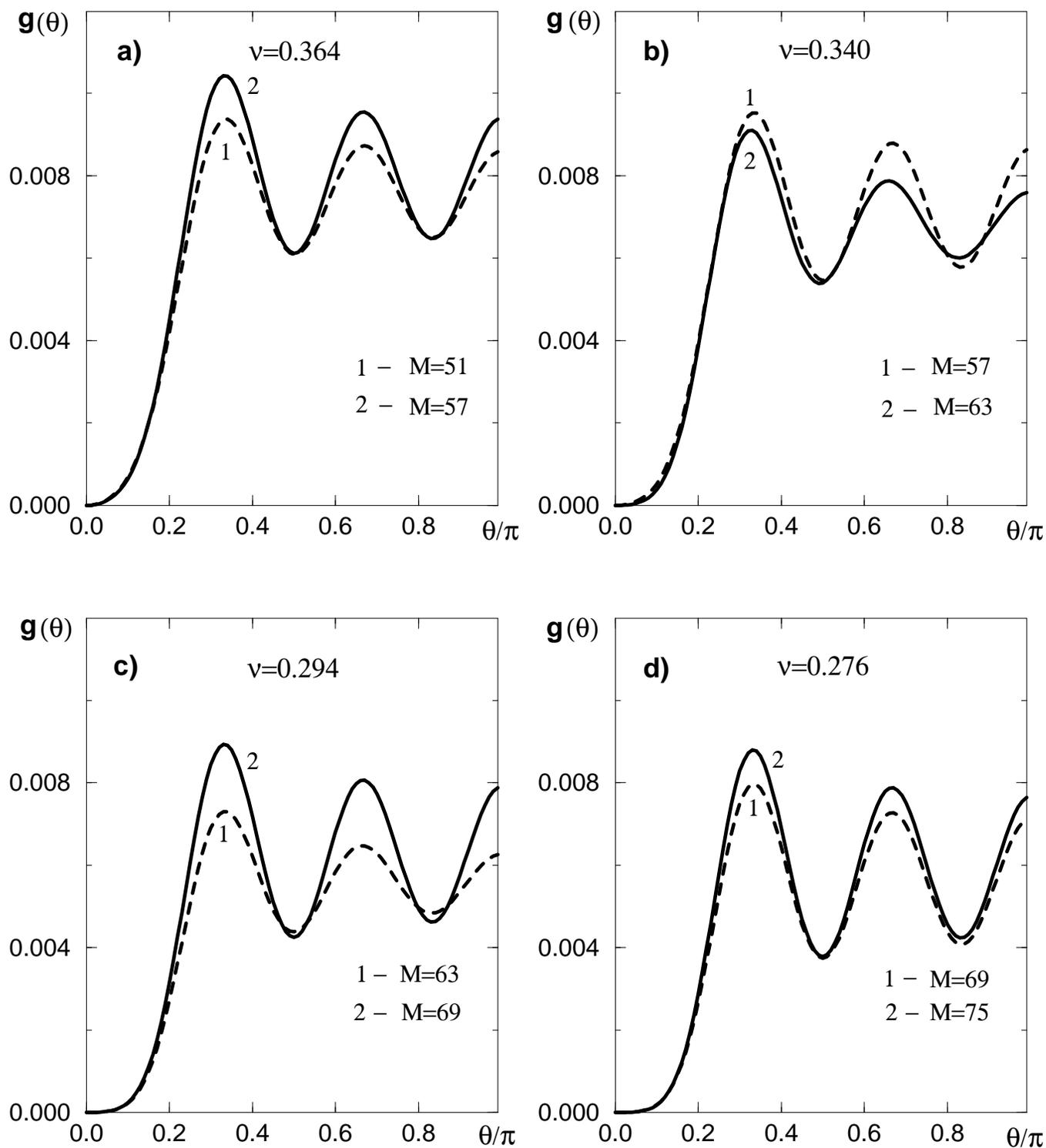
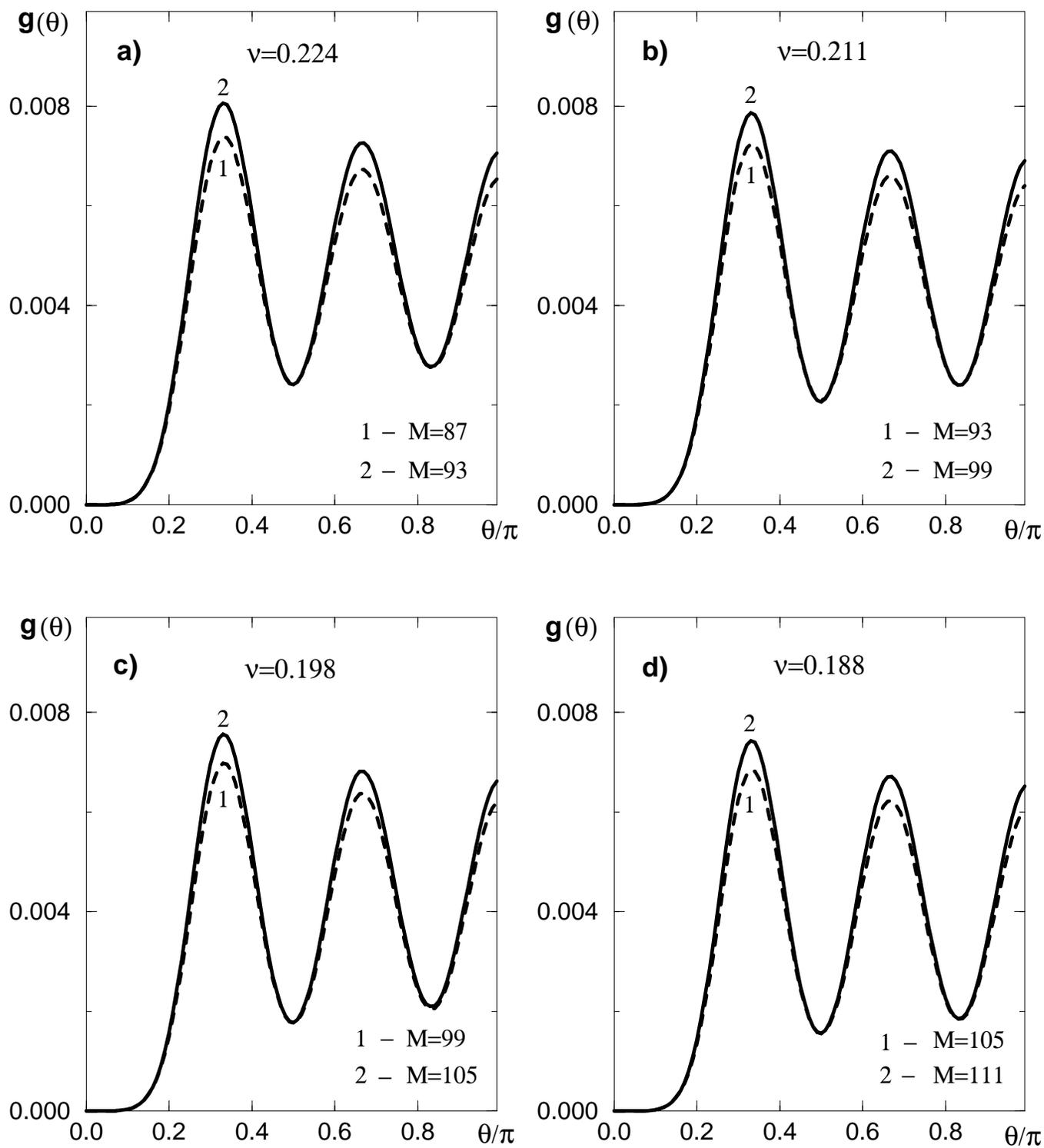
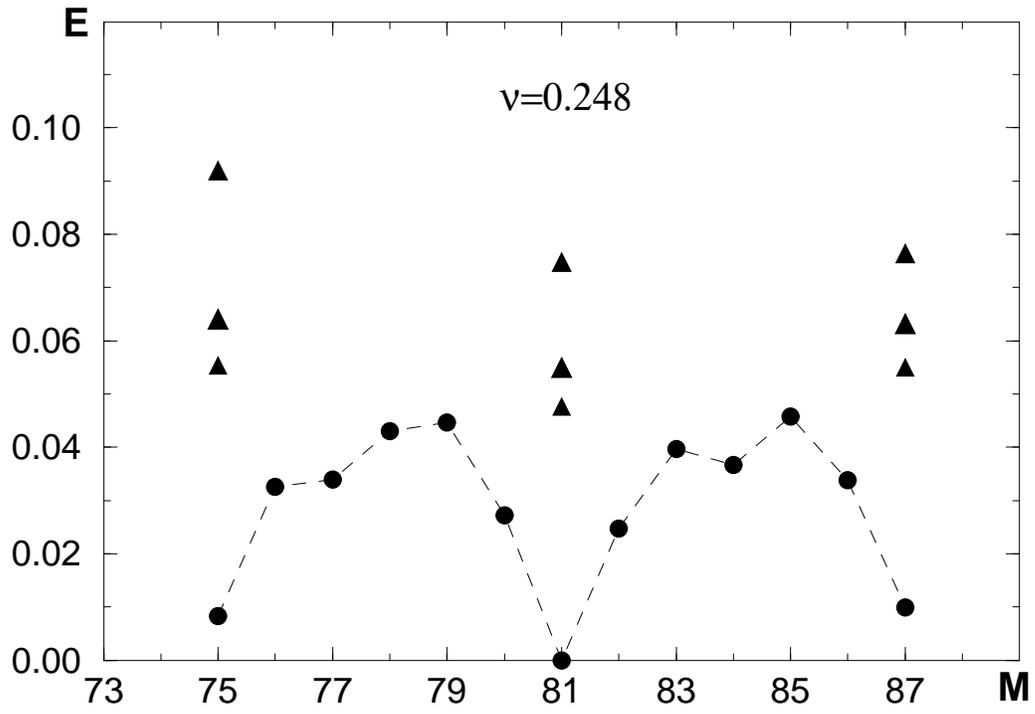
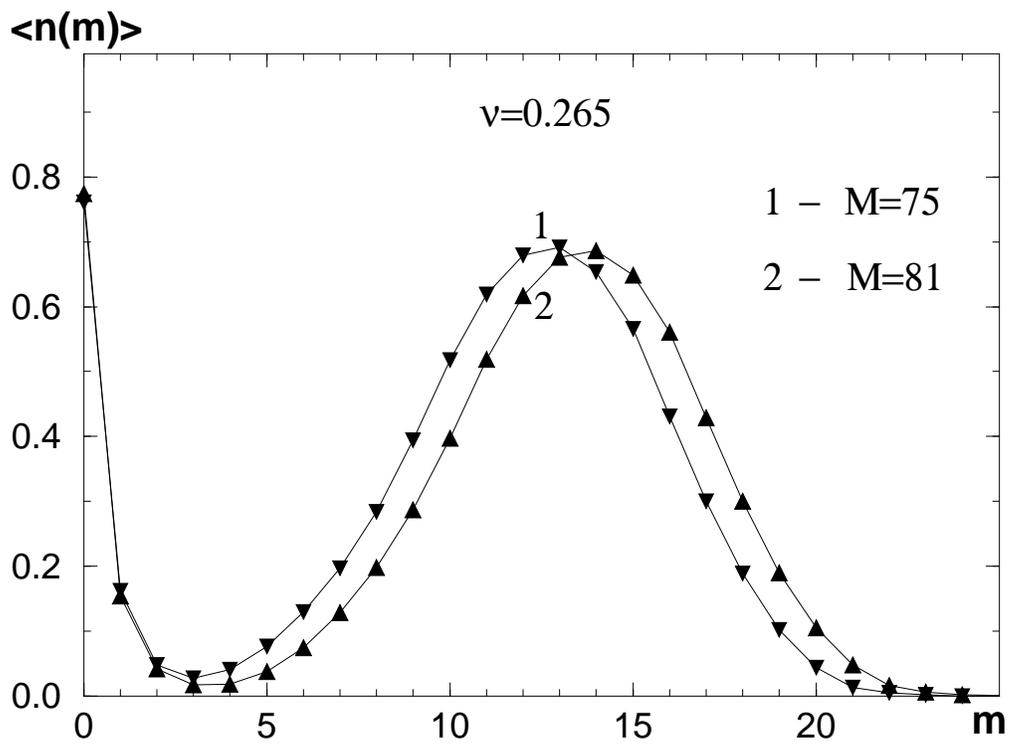
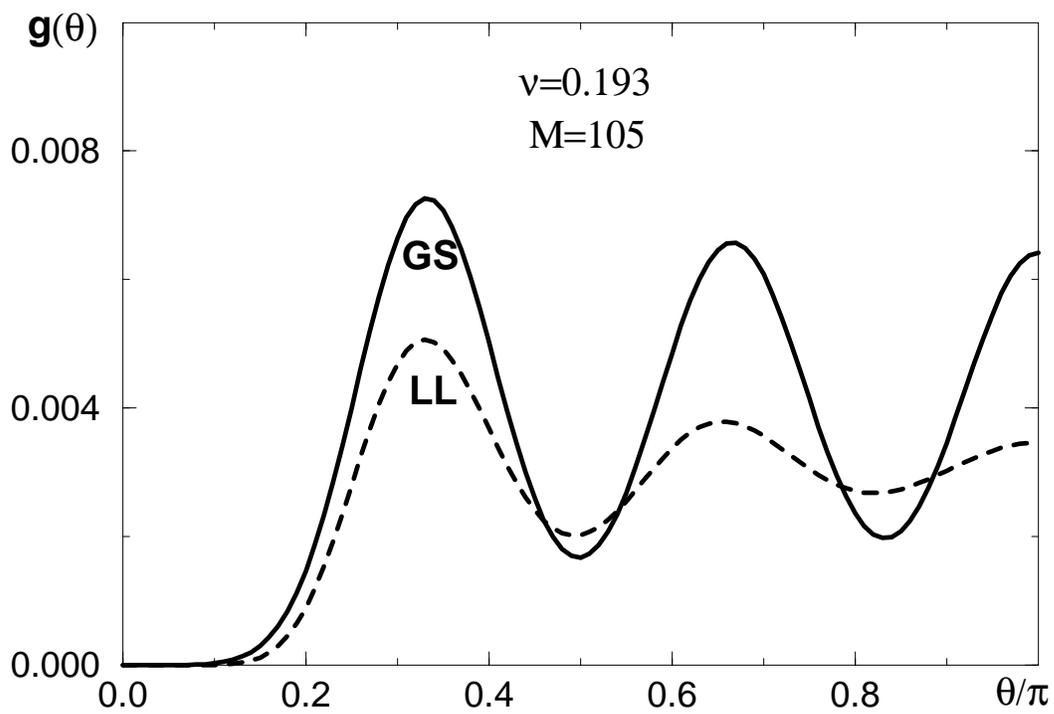
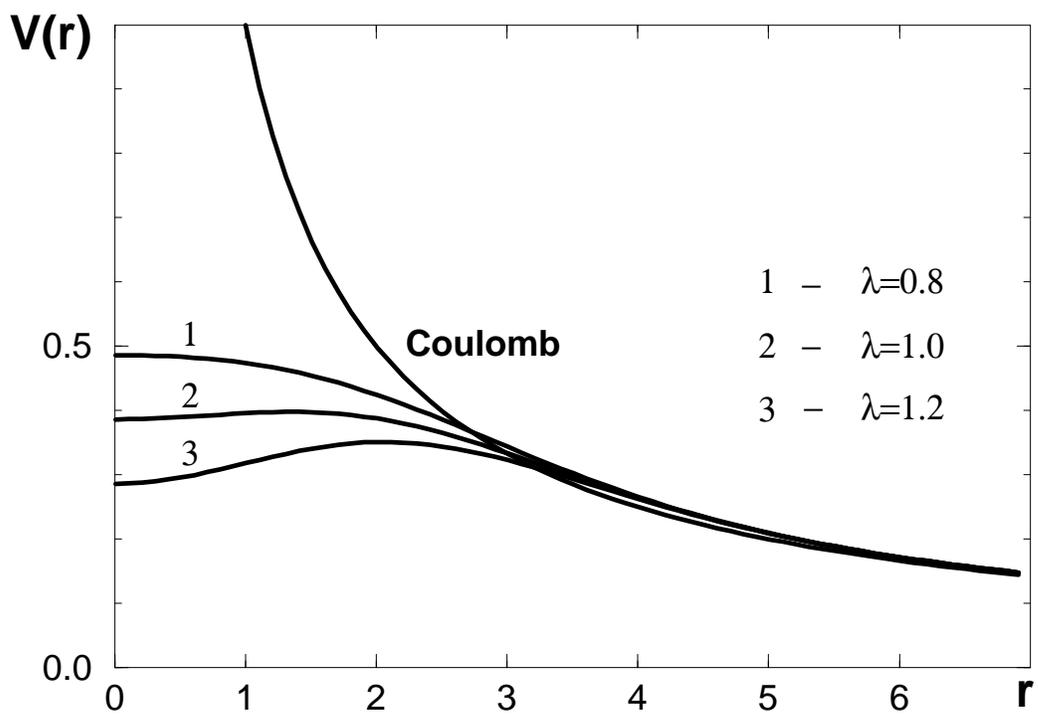
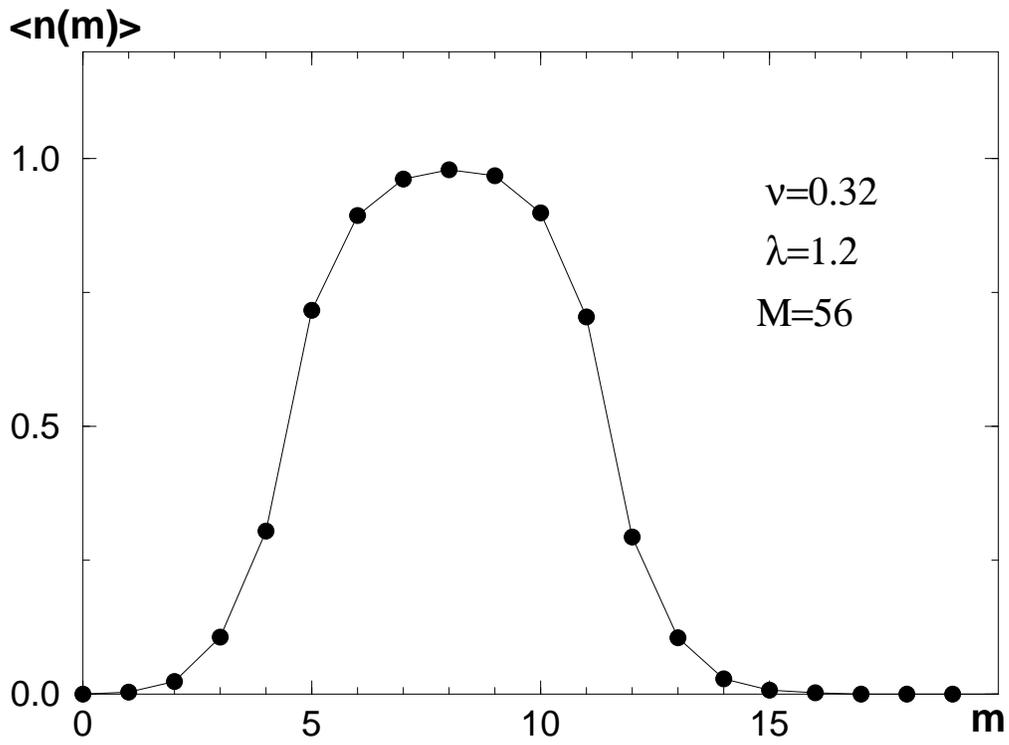
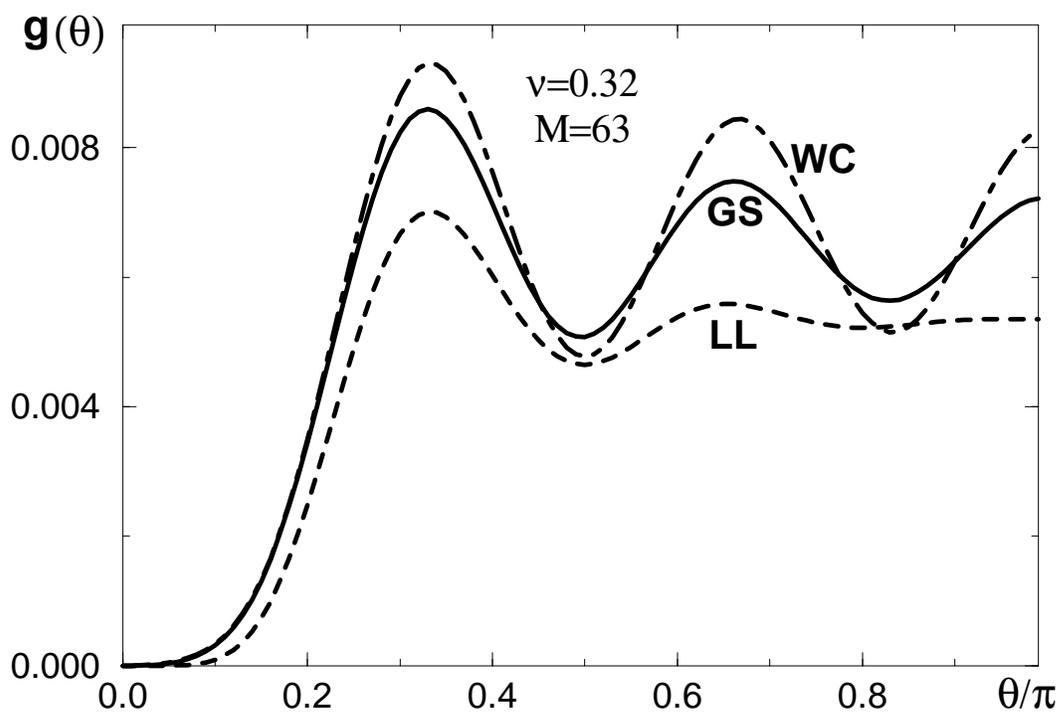
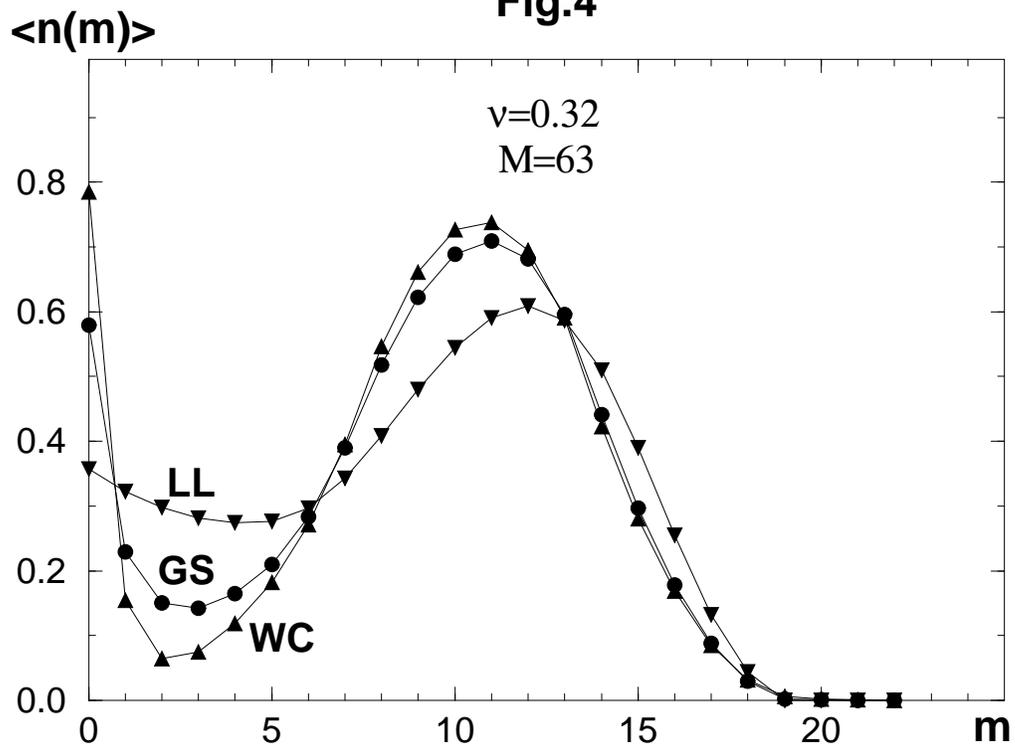
\end{document}